\documentstyle[preprint,aps] {revtex}                 

\begin{document}

\draft
\title{Test of low--energy theorems for $\mathbf{p(\vec{\gamma},\pi^{0})p}$ in the threshold region}

\author{A.~Schmidt$^1$, P.~Achenbach$^3$, J.~Ahrens$^1$, H.~J.~Arends$^1$, R.~Beck$^1$\footnote[1]{corresponding author, email: rbeck@kph.uni-mainz.de},
A.~M.~Bernstein$^2$, V.~Hejny$^3$,
M.~Kotulla$^4$, B.~Krusche$^5$, V.~Kuhr$^6$, R.~Leukel$^1$, I.~J.~D.~MacGregor$^7$, J.~C.~McGeorge$^7$, 
V.~Metag$^4$, V.~M.~Olmos de Le\'on$^1$, F.~Rambo$^6$, U.~Siodlaczek$^8$,
H.~Str{\"o}her$^3$, Th.~Walcher$^1$, J.~Wei{\ss}$^4$, F.~Wissmann$^6$ and M.~Wolf$^4$}
\address{$^1$Institut f\"ur Kernphysik, Universit\"at Mainz, 55099 Mainz, Germany\\
$^2$Departement of Physics and Laboratory for Nuclear Science, MIT, Boston, MA, USA\\
$^3$Institut f\"ur Kernphysik, Forschungszentrum J\"ulich GmbH, 52425 J\"ulich, Germany\\
$^4$II.~Physikalisches Institut, Justus--Liebig--Universit\"at Gie{\ss}en,
35392 Gie{\ss}en, Germany\\
$^5$Department f\"ur Physik und Astronomie, Universit\"at Basel, 4056 Basel, Switzerland\\
$^6$II.~Physikalisches Institut, Georg--August--Universit\"at G\"ottingen, 37073 G\"ottingen, Germany\\
$^7$Department of Physics and Astronomy, Glasgow University, Glasgow G128QQ, UK\\
$^8$Physikalisches Institut, Eberhard-Karls-Universit\"at T\"ubingen, 72076 T\"ubingen, Germany}

\date{\today}

\maketitle

\begin{abstract}
The photon asymmetry in the reaction $p(\vec{\gamma},\pi^{0})p$ 
close to threshold has been measured for the first time with the photon spectrometer TAPS
using linearly polarized photons from the tagged--photon facility at the Mainz
Microtron MAMI. 
The total and differential cross sections were also measured simultaneously with the photon asymmetry.
This allowed determination of the $S$--wave and all three $P$-wave amplitudes. 
The values obtained at threshold are
$E_{0+} = (-1.33 \pm 0.08_{stat} \pm 0.03_{sys})10^{-3}/m_{\pi^+}$, 
$P_1 = (9.47 \pm 0.08_{stat} \pm 0.29_{sys}) 10^{-3} q/m^2_{\pi^+}$,
$P_2 = (-9.46 \pm 0.1_{stat} \pm 0.29_{sys}) 10^{-3} q/m^2_{\pi^+}$ and
$P_3 = (11.48 \pm 0.06_{stat} \pm 0.35_{sys}) 10^{-3} q/m^2_{\pi^+}$. 
The low--energy theorems based on the parameter--free third--order calculations of heavy--baryon 
chiral perturbation theory for $P_1$ and $P_2$ agree with the 
experimental values.
\end{abstract}

In the early 70's, low--energy theorems (LETs) were derived for the amplitudes
of pion photoproduction from the nucleon at threshold \cite{Bae70,Vai72}. 
Based on fundamental principles, 
like gauge invariance and the partially conserved axial current, the LETs predict the 
value of the $S$--wave threshold amplitude $E_{0+}$ in a power series 
in $\mu=m_\pi/m_N$, the ratio of the masses of the pion and nucleon. 
The LETs represent tests of effective degrees of freedom in the non--perturbative domain of QCD 
and, therefore, their investigation is of considerable interest for an understanding
of QCD at low momentum transfers. 
Only the development of high duty factor accelerators enabled first precise measurements 
of the photoproduction of neutral pions from the proton at Saclay~\cite{Maz86} and Mainz~\cite{Bec90}.
The experimental values for $E_{0+}$ at threshold were in conflict with the LET prediction.
Most calculations also failed to predict the strong dependence of $E_{0+}$  
on the photon energy between the $\pi^0$--threshold (144.7 MeV) and 160 MeV, where a unitary cusp 
due to the two--step process $\gamma p \rightarrow \pi^+ n \rightarrow \pi^0 p$~\cite{Fae80} 
was seen in the Mainz measurement~\cite{Bec90}.
These disagreements motivated several theoretical and experimental investigations. New  
experiments were performed at Mainz~\cite{Fuc96} and Saskatoon~\cite{Berg96}, 
measuring the total and differential cross sections close to threshold. The 
extracted values of $E_{0+}$ confirmed the strong energy dependence and were again 
nearly a factor of two smaller than the LET prediction at threshold. This discrepancy was explained
by Bernard, Kaiser and Mei{\ss}ner~\cite{Ber91}, who investigated threshold pion photoproduction in 
the framework of heavy--baryon chiral perturbation 
theory (ChPT), which showed that additional contributions due to pion loops in $\mu^2$ have to be added 
to the old LET.

In the following years, refined calculations within heavy--baryon ChPT~\cite{Ber96a} 
led to descriptions of the four relevant amplitudes at threshold by well--defined 
expansions up to order 
$p^4$ in the $S$--wave amplitude $E_{0+}$ and $p^3$ in the $P$--wave 
combinations $P_1$, $P_2$ and $P_3$, where $p$ denotes any small momentum or pion mass, 
the expansion parameters in heavy--baryon ChPT.
To that order, three low--energy constants (LEC) due to the renormalization counter 
terms appear, two in the expansion of $E_{0+}$ and an additional LEC $b_P$ for $P_3$, 
which have to be fitted to the data or estimated by resonance saturation.
However, two combinations of the $P$--wave amplitudes, $P_1$ and $P_2$, are free of low--energy constants.
Their expansions in $\mu$ converge rather well leading to new LETs for these combinations.
Therefore, the P--wave LETs offer a significant test of heavy--baryon ChPT.

However, for this test the S--wave amplitude $E_{0+}$ and the three P--wave combinations
$P_1$, $P_2$ and $P_3$ have to be separated. This separation can be achieved by measuring
the photon asymmetry using linearly polarized photons, in addition to the measurement of the total and 
differential cross sections.
The $p(\vec{\gamma},\pi^{0})p$ experiment~\cite{Sch01}, reported in this letter, 
was performed at the Mainz Microtron MAMI~\cite{Her90}
using the Glasgow/Mainz tagged photon facility~\cite{Ant91,Hal95} and the photon spectrometer
TAPS~\cite{Nov91}.
The MAMI accelerator delivered a continuous wave beam of 405 MeV electrons.
Linearly polarized photons were produced via coherent bremsstrahlung 
in a 100 $\mu m$ thick diamond radiator~\cite{Loh94,Sch95} with degrees of polarization of up to 50\%.
The diamond radiator was mounted on a goniometer~\cite{Sch95}, which was adjusted so that 
the linearly polarized photons lay in the energy region between $\pi^0$--threshold and 166 MeV.
The energy of the photons was determined by measuring the energy of the electron after the
bremsstrahlung process with the tagging spectrometer. The resolution was approximately \mbox{1 MeV} at 
intensities of up to 5$\times 10^5$ photons s$^{-1}$ MeV$^{-1}$.

Neutral pions were produced in a liquid hydrogen target of 
cylindrical shape with a length of 10 cm and a diameter of 4 cm. 
The neutral pion decay photons were detected in TAPS, consisting of six blocks of 
hexagonally shaped $\rm BaF_2$ scintillation crystals each arranged in a matrix of $8\times8$ 
detectors. The blocks were mounted in a horizontal plane around the target at polar angles of $\pm50$, 
$\pm100$, and $\pm150$ degrees with respect to the photon beam direction. A forward wall, 
consisting of 120 phoswich telescopes~\cite{Nov96}, covered polar angles between 5 and 20 degrees.
Further details of the experimental set--up are found in Ref.~\cite{Hej00}.
 
The identification of neutral pions relies on the coincident detection of the two photons from 
$\pi^0$--decay in the TAPS detector (the $\pi^0 \rightarrow \gamma\gamma$ branching 
ratio is $\approx$ 99.8\%).
The photons were identified with the help of charged--particle veto detectors, a pulse shape and a 
time--of--flight analysis. An invariant mass analysis was performed to identify neutral pions
and a resolution of $\simeq$ 19~MeV (FWHM) was achieved. 
Accidental coincidences between TAPS and the tagging spectrometer were subtracted using 
scaled distributions of background events outside the prompt coincidence time window. 
For each event a missing energy analysis 
was performed for an unambiguous identification of neutral pions in the threshold region. 
The missing energy resolution for $\pi^0$--mesons close to threshold was approximately 5 MeV (FWHM).
The acceptance of TAPS for neutral pions and the analysing efficiency were determined 
by a Monte Carlo simulation using the GEANT3 code~\cite{Bru86} in which all relevant properties 
of the setup and the TAPS detectors were taken into account.

The differential cross sections can be expressed in terms of the $S$-- and $P$--wave multipoles, 
assuming that close to threshold neutral pions are only produced with angular momenta $l_\pi$ 
of zero and one. Due to parity and angular momentum conservation only the $S$--wave 
amplitude $E_{0+}$ ($l_\pi=0$)
and the $P$--wave amplitudes $M_{1+}$, $M_{1-}$ and $E_{1+}$ ($l_\pi=1$) 
can contribute and it is convenient to 
write the differential cross section and the photon asymmetry in terms of the three $P$--wave 
combinations 
$P_1=3E_{1+}+M_{1+}-M_{1-}$, $P_2=3E_{1+}-M_{1+}+M_{1-}$ and 
$P_3=2M_{1+}+M_{1-}$.
The c.m.~differential cross section is
\begin{equation}
\frac{d\sigma(\theta)}{d\Omega} = \frac{q}{k}(A+B~cos(\theta)+C~cos^2(\theta))\,\,,
\end{equation}
where $\theta$ is the c.m.~polar angle of the pion with respect to the beam direction and 
$q$ and $k$ denote the c.m.~momenta of pion and photon, respectively. 
The coefficients $A=|E_{0+}|^2+|P_{23}|^2$, $B=2Re(E_{0+}P_1^\ast)$
and $C=|P_1|^2-|P_{23}|^2$ are functions of the multipole amplitudes with 
$P_{23}^{\,2}=\frac{1}{2}(P_2^{\,2}+P_3^{\,2})$.
Earlier measurements of the total and differential cross sections already allowed 
determination of $E_{0+}$, $P_1$ and the combination $P_{23}$. 
In order to obtain $E_{0+}$ and all three $P$--waves separately, it is necessary to measure, in 
addition to the cross sections, the photon asymmetry $\Sigma$,
\begin{equation}
\Sigma = \frac{d\sigma_{\perp} -d\sigma_{\parallel}}
{d\sigma_{\perp} + d\sigma_{\parallel}}\,\,,
\end{equation}
where $d\sigma_{\perp}$ and $d\sigma_{\parallel}$ are the differential cross sections for photon 
polarizations perpendicular and parallel to the reaction plane defined by the pion and proton. 
The asymmetry is proportional to the difference of the squares of $P_3$ and $P_2$:
\begin{equation}
\Sigma(\theta)=\frac{q}{2k}(P_3^2-P_2^2)\cdot sin^2(\theta)/\frac{d\sigma(\theta)}{d\Omega}\,.
\end{equation}
Thus, the measurement of the total and differential cross sections together with 
$\Sigma$ allows a separate determination of $P_2$ and $P_3$ and hence a test of the 
new LET of ChPT~\cite{Ber96a}.

\begin{figure}
\centerline{\psfig{figure=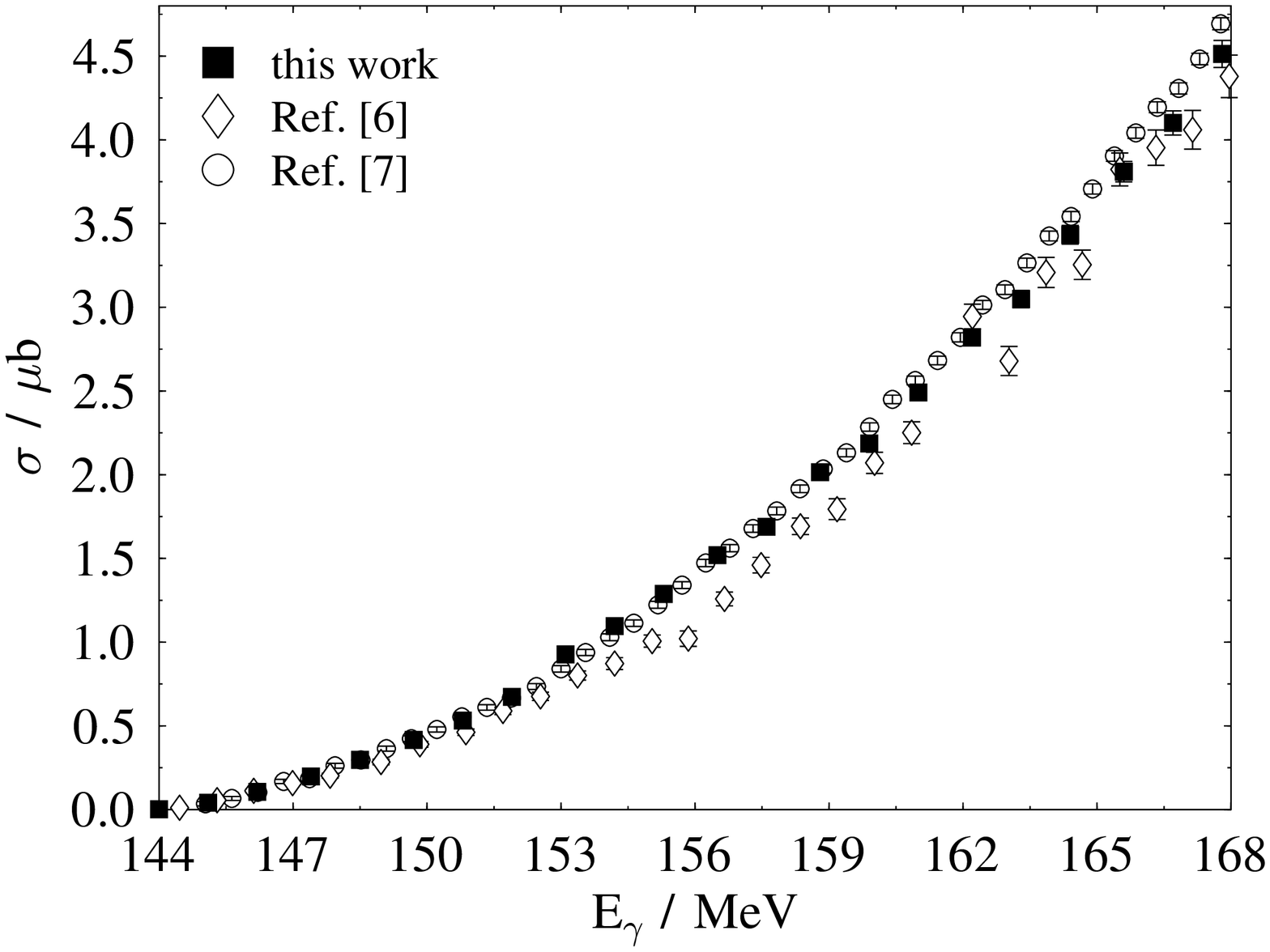,width=10cm}}
\caption{Total cross sections for $\pi^0$ photoproduction close to threshold with statistical errors 
(without systematic error of 5\%) as function of incident photon energy (solid squares, this work, 
open circles, Ref. \protect\cite{Berg96}, open diamonds Ref. \protect\cite{Fuc96}).}
\label{totwq}
\end{figure}

\begin{figure}
\centerline{\psfig{figure=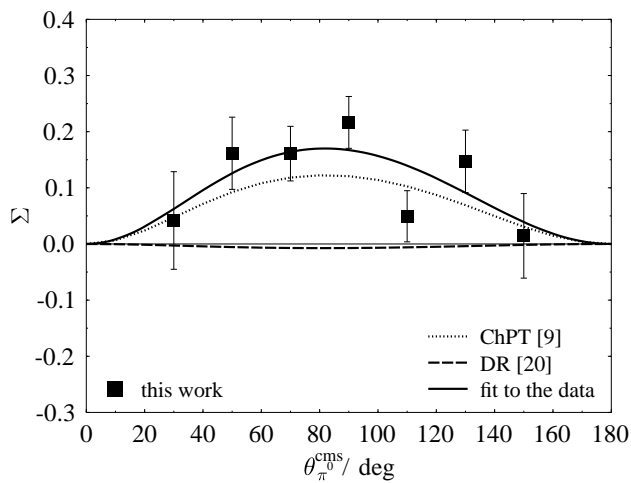,width=10cm}}
\caption{Photon asymmetry for $\pi^0$ photoproduction for a photon energy of 159.5 MeV with 
statistical errors (without systematic error of 3\%) as a function of the polar angle $\theta$ 
(solid line: fit to the data) in comparison to ChPT~\protect\cite{Ber96a} 
(dotted line) and DR~\protect\cite{Han97} (dashed line). With the new value of the low--energy constant
$b_P$ the ChPT calculation~\protect\cite{Ber01} is in agreement 
with the experimental values.}
\label{asym}
\end{figure}

In the present work the total and differential cross sections were measured over 
the energy range from
$\pi^0$--threshold to 168 MeV. Fig.~\ref{totwq} shows the results for the total 
cross section which agrees with the data of Ref.~\cite{Berg96}; the results of 
Ref.~\cite{Fuc96} are systematically lower, at least in the incident photon energy range of
153--162 MeV. This discrepancy may be due to a better elimination of pions produced in the 
target cell windows, performed in the analysis of the present data,
combined with the improved detector acceptance for forward and backward angles.
The different slope in the total cross section of~\cite{Fuc96} compared to the other 
experiments results in a steeper energy dependence for the 
real part of $E_{0+}$ and slightly smaller values for $P_1$ and $P_{23}$ (see Table~\ref{mult}). 
The results for the photon asymmetry are shown in Fig.~\ref{asym} 
in comparison to the values of ChPT~\cite{Ber96a} and to a prediction of a dispersion theoretical
calculation (DR) by Hanstein, Drechsel and Tiator~\cite{Han97}. 
The photon asymmetry was determined from all the data between threshold and 166 MeV for which the 
mean energy was 159.5 MeV. The theoretical predictions are shown for the same energy.
The energy dependence of the ChPT prediction for the photon asymmetry was
explored in the range threshold to 166 MeV and found to have a very
small effect on the average, eg. $<\,2\%$ at $90^\circ$. 

The values for the real and imaginary part of $E_{0+}$ and the three $P$--wave combinations were 
extracted via two multipole fits to the cross sections and the photon asymmetry simultaneously
using the following minimal model assumptions.
The parameterizations of $ReE_{0+}$ and $ImE_{0+}$ take into account the strong energy 
dependence of $E_{0+}$ below and above $\pi^+$--threshold ($E_{thr}^{n\pi^+}=$151.4 MeV) due to the 
two--step process $\gamma p \rightarrow \pi^+ n \rightarrow \pi^0 p$~\cite{Bern97}:
\begin{equation}
E_{0+}(E_\gamma)=A^{p\pi^0}(E_\gamma)+i\,\beta\,q_{\pi^+}\,,
\end{equation}
where $q_{\pi^+}$ is the $\pi^+$ c.m.~momentum.
$E_{0+}$ is a sum of two parts, $A^{p\pi^0}$ due to the direct process and a second part, 
arising from the two--step process.
Below $\pi^+$--threshold, one must analytically continue $q_{\pi^+}\rightarrow i|q_{\pi^+}|$. Thus 
$E_{0+}$ is purely real and has the value $E_{0+}=A^{p\pi^0}-\beta |q_{\pi^+}|$, where 
$\beta$ is the product 
of the S--wave amplitude $E_{0+}^{n\pi^+}$ for $\pi^+$--production and the scattering 
length $a_{n\pi^+ \rightarrow p\pi^0}$. Above $\pi^+$--threshold, $E_{0+}$ is complex with 
$E_{0+}=A^{p\pi^0}+i\,\beta |q_{\pi^+}|$ and $ImE_{0+}=\,\beta |q_{\pi^+}|$, the cusp function.
In the threshold region the imaginary parts of the P-waves are negligible because of the small 
$\pi N$--phase shifts.
The two multipole fits differ in the energy dependence of the real parts of the P--wave combinations.
For the first fit the usual assumption of a behaviour proportional 
to the product of $q$ and $k$ was adopted ($q k$--fit, $\chi^2/dof=1.28$).
The assumption made for the second fit is an energy dependence of the P--wave amplitudes 
proportional to $q$ ($q$--fit, $\chi^2/dof=1.29$). This is the dependence which ChPT predicts for 
the P--wave amplitudes in the near--threshold region, but at higher energies the prediction 
is in between the $q$ and $qk$ energy dependence.

\begin{figure}
\centerline{\psfig{figure=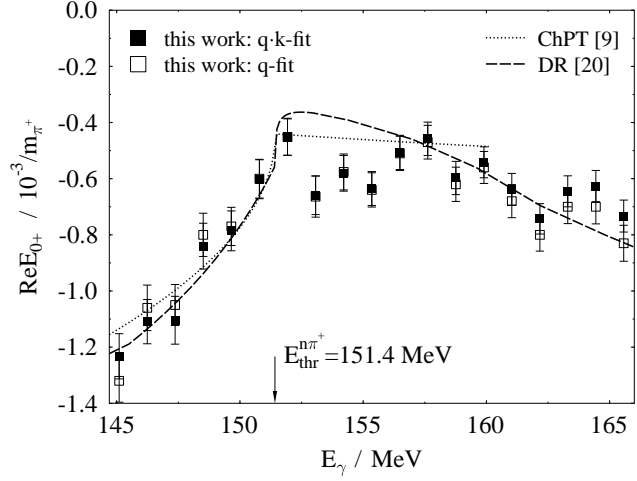,width=10cm}}
\caption{Results for $ReE_{0+}$ with statistical~errors as a function of incident photon energy $E_\gamma$
for an assumed energy dependence of the P--wave amplitudes
proportional to $q\cdot k$ (solid squares) and $q$ (open squares)    
in comparison to ChPT~\protect\cite{Ber96a} (dotted line) and DR~\protect\cite{Han97} (dashed line).}
\label{reeop}
\end{figure}

The results of both multipole fits for $ReE_{0+}$ as a function of the incident photon energy
are shown in Fig.~\ref{reeop} and compared with the predictions of ChPT 
and of DR. The results for the threshold values of $ReE_{0+}$ (at the $\pi^0$-- and $\pi^+$--threshold), 
for the parameter $\beta$ of $ImE_{0+}$ and for the values of the threshold slopes of the three P--wave
combinations
of the $q k$--fit and the $q$--fit are summarized in Table~\ref{mult} together with the results of 
~\cite{Berg96} and ~\cite{Fuc96,Bern97}. To obtain the threshold slope of the $qk$--fits the values of the 
P--wave combinations of these fits (unit: $q k\cdot 10^{-3}/m_{\pi^+}^3$) must be multiplied by the 
threshold value of the photon momentum $k$. 
In addition the results are compared to the ChPT and DR predictions, where the errors of ChPT refer to a 
5\% theoretical uncertainty. 

The extracted value for $\beta$ and thus
$ImE_{0+}$ of the $q$--fit is larger than the value of $\beta$ obtained with the $q k$--fit.
This result can be explained by the observation that A
is the best measured of the three coefficients of the differential cross section, and by noting that this 
determines the absolute value of $E_{0+}$ in addition to the dominant P--wave contribution.
Since $ReE_{0+}$ is determined from the B coefficient this gives $ImE_{0+}$ after a subtraction of
the P--wave contribution to the A coefficient. If one assumes a smaller energy dependence in the 
P--wave amplitudes ($q$--fit), a stronger energy dependence for $ImE_{0+}$ will result.
However, the values of both fits for $ReE_{0+}$ and the values of the three P--wave combinations 
at threshold are in remarkable agreement.
Assuming for $E_{0+}^{n\pi^+}$ the prediction of ChPT, which agrees with the result of ~\cite{Kor99},
taking for $a_{n\pi^+ \rightarrow p\pi^0}$ the measured value of~\cite{Schr99} and thus fixing the 
parameter $\beta$ to the expected unitary value of $3.61\cdot10^{-3}/m_{\pi^+}^2$, the values 
of the P--wave combinations for both fits change by less than 3\%.

The main experimental uncertainty is the value of $\beta$. The systematic error of $\beta$ 
in Table~\ref{mult} includes the experimental uncertainty in the energy 
dependence of the P--wave amplitudes. The average value for 
$\beta = (3.8 \pm 1.4)\cdot 10^{-3}/m_{\pi^+}^2$
of the two fit results, obtained in this experiment, is consistent with the unitary
value. To determine $\beta$ more accurately will require a 
direct measurement of $ImE_{0+}$ in the $\vec{p}(\gamma,\pi^{0})p$ reaction with a
polarized target~\cite{Bern98}.

For both fits the low--energy theorems of ChPT (${\cal{O}}(p^3)$) for $P_1$ and $P_2$ agree with the 
measured experimental results within their systematic and statistical errors.
The experimental value for $P_3$ is higher than the value of
ChPT, which can be explained by the smaller total and differential cross sections of Ref.~\cite{Fuc96}, 
used by ChPT to determine the dominant low--energy constant $b_P$ for this multipole. 

A new fourth--order calculation in heavy--baryon ChPT by Bernard et al., introduced in \cite{Ber01}
and compared to the new MAMI data presented in this letter, shows, that the 
potentially large $\Delta$--isobar contributions are cancelled by the fourth--order loop 
corrections to the P--wave low--energy theorems. 
This gives confidence in the third--order 
LET predictions for $P_1$ and $P_2$, which are in agreement with the present MAMI data.
With the new value of $b_P$~\cite{Ber01}, fitted to the present MAMI data, the ChPT calculation is in 
agreement with the measured photon asymmetry.

To summarize, the total and differential cross sections and the photon asymmetry
for the reaction $p(\vec{\gamma},\pi^{0})p$
have been measured simultaneously for the first time in the threshold region.
Using a multipole fit to the physical observables the threshold values of the $S$--wave amplitude $E_{0+}$ 
and all three $P$--wave amplitudes were extracted.
The main conclusion is that the calculations of heavy--baryon ChPT for $P_1$ and $P_2$
are in agreement with the experimental results.

The authors wish to acknowledge the excellent support of K.H. Kaiser, H. Euteneuer
and the accelerator group of MAMI, as well as many other scientists and
technicians of the Institut f\"{u}r Kernphysik at Mainz. We would like to thank  also 
D. Drechsel, O. Hanstein, L. Tiator and U. Mei{\ss}ner for very fruitful discussions and comments.
A.M. Bernstein is grateful to the Alexander von Humboldt Foundation for a Research Award.
This work was supported by the Deutsche Forschungsgemeinschaft (SFB~443) and
the UK Engineering and Physical Sciences Research Council.

\input{psfig}

\input{psfig}

\input{psfig}

\begin{table}[h]\centering
\begin{tabular}{c|c|c|c|c|c|c}
&\multicolumn{2}{c|}{this work} & Bergstrom$^a$ & Fuchs$^a$ & ChPT & DR$^a$\\
\hline
& $qk$--fit$^a$ & $q$--fit & $qk$--fit & $qk$--fit & & \\
\hline
$ E_{0+}(E_{thr}^{p\pi^0})$ & $-1.23\pm0.08\pm0.03$ & $-1.33\pm0.08\pm0.03$ & $ -1.32\pm0.05$ & $-1.31\pm 0.2$  & -1.16 & -1.22\\
$ E_{0+}(E_{thr}^{n\pi^+})$ & $-0.45\pm0.07\pm0.02 $ &$-0.45\pm0.06\pm0.02 $ &$ -0.52\pm0.04$ &$-0.34\pm 0.03$ & -0.43 & -0.56\\
$\beta$ & $2.43\pm0.28\pm1.0$ & $5.2\pm0.2\pm1.0$ & 3.0--3.8 & $2.82\pm0.32$ & 2.78& 3.6\\
$ {P_1}$ & $ 9.46\pm0.05\pm0.28$ &$9.47\pm0.08\pm0.29$ & $ 9.3\pm0.09$ & $ 9.08\pm 0.14$ & $9.14\pm0.5$& 9.55\\
$ {P_2}$ & $-9.5\pm0.09\pm0.28 $&$ -9.46\pm0.1\pm0.29 $ &  &  &  $-9.7\pm0.5$ & -10.37\\ 
$ {P_3}$ & $11.32\pm0.11\pm0.34$ &$11.48\pm0.06\pm0.35$ &  &  & $10.36$ & 9.27\\
$ P_{23}$ & $ 10.45\pm 0.07$ &$ 10.52\pm0.06$ & $10.53\pm 0.07$ &  $10.37\pm0.08$ &  11.07 & 9.84\\
\end{tabular}
\caption{Results of both fits ($qk$--fit and $q$--fit) for $ReE_{0+}$ at the $\pi^0$-- and 
$\pi^+$--threshold (unit: $10^{-3}/m_{\pi^+}$), for the parameter $\beta$ of $ImE_{0+}$ 
(unit: $10^{-3}/m_{\pi^+}^2$) and for the three combinations of the $P$--wave amplitudes 
(unit: $q \cdot 10^{-3}/m_{\pi^+}^2$) 
with statistical and systematic~errors in comparison to the results of previous 
experiments (~\protect\cite{Berg96} 
and ~\protect\cite{Fuc96,Bern97}, only with statistical errors) and to the predictions of 
ChPT~\protect\cite{Ber96a,Ber96b} (${\cal{O}}$($p^3$)) and of a dispersion theoretical 
approach (DR,~\protect\cite{Han97}). ($^a$\, Values of the P--wave combinations converted 
into the unit $q \cdot 10^{-3}/m_{\pi^+}^2$.)}
\label{mult}
\end{table}

\end{document}